# A Magnetic and Mössbauer Spectral Study of Core/Shell Structured Fe/Au Nanoparticles

Sung-Jin Cho[1], Ahmed M. Shahin[2], Gary J. Long[2*], Joseph E. Davies[3], Kai Liu[3*], Fernande Grandjean[4], and Susan M. Kauzlarich[1*]

[1]*Department of Chemistry, University of California, Davis, CA 95616*
[2]*Department of Chemistry, University of Missouri-Rolla, Rolla, MO 65409-0010*
[3]*Department of Physics, University of California, Davis, CA 95616*
[4]*Department of Physics, University of Liège, B5, B-4000 Sart-Tilman, Belgium*

(12/16/2005)

## Abstract

Fe/Au nanoparticles have been chemically synthesized through a reverse micelle reaction and investigated by both conventional and synchrotron based x-ray diffraction and by magnetic and Mössbauer spectral studies. The powder x-ray diffraction patterns reveal both the presence of crystalline α-iron and gold and the absence of any crystalline iron oxides or other crystalline products. First-order reversal curves, along with the major hysteresis loops of the Fe/Au nanoparticles have been measured as a function of time in order to investigate the evolution of their magnetic properties. The iron-57 Mössbauer spectra of both uncoated iron nanoparticles and the Fe/Au nanoparticles have been measured at 78 and 295 K and indicate that two major iron containing components are present, namely the expected α-iron and the unexpected amorphous $Fe_{1-x}B_x$ alloy; several poorly crystallized ordered iron(III) oxide components as well as paramagnetic iron(II) and iron(III) components are also observed. These results indicate that the Fe-core/Au-shell nanoparticles synthesized through reverse micelles are far more complex that had been believed.

*Corresponding Authors: glong@umr.edu, kailiu@ucdavis.edu, smkauzlarich@ucdavis.edu

**Introduction**

As a technique for the magnetic characterization and structural analysis of solids, Mössbauer spectroscopy has provided important microscopic information for over four decades.[1,2] For the past 20 years, it has contributed to our understanding of the properties of new nanostructured materials.[3] Nanostructured materials can range from crystalline to amorphous materials, and their physical properties have been investigated in detail[4-7] because the nanostructured state can lead to useful differences in the structural, magnetic, electronic, and chemical properties of materials.[4-7] Perhaps the major interest in recent magnetic and Mössbauer spectral studies of nanostructured materials lies less with materials that simply have nanoscale sizes but, rather, with materials that have purposely been synthesized to have a unique nanostructure.[8-11] This type of study is well illustrated by the work of Herr et al.[12] on the 77 K spectrum of nanostructured iron.

Glavee et al.[13] have investigated the formation of nanoscale iron obtained from the borohydride reduction of aqueous iron(II) and iron(III) in non-aqueous media with Mössbauer spectroscopy; the analysis of the Mössbauer spectra confirmed the presence of $\alpha$-iron in their material together with several other iron containing phases, such as iron-boron alloys, $Fe_2B$, and $Fe_3B$.

Recently there has been interest in gold-coated magnetic nanoparticles because of their ability to yield high magnetic moments with the simplicity of bioconjugation to the gold surface. More specifically, both nanoparticles of $\gamma$-$Fe_2O_3$ with a gold shell[14] and gold coated cobalt nanoparticles[15] have been studied.

Initial studies on core/shell structured Fe/Au nanoparticles have been reported[16] and a subsequent x-ray absorption study[17] has shown that such particles often have cores containing oxidized iron. The oxidation state of the core iron based material has been of key interest in this research because it will determine the magnetic properties of the core/shell nanostructured material.[18] If iron nanoparticle cores can be protected from oxidation, Fe/Au nanoparticles have the potential to provide a large magnetic moment in a small diameter particle. The gold coating was originally envisioned to both protect the iron core from oxidation and to provide a platform for bioconjugation. We have shown that the surface coating is rough and have postulated that the Fe/Au nanoparticles oxidize overtime unless they are further protected.[18, 19] We have previously characterized[18, 19] these nanoparticles by x-ray diffraction, transmission electron microscopy, and magnetometry. This paper reports synchrotron x-ray diffraction, magnetic, and Mössbauer spectral studies of these materials.

**Experimental**

All preparative reactions were carried out in a 250 mL three-necked round-bottom flask that was attached to a Schlenk line. All chemicals were used as received without further purification and water, obtained with a Barnstead ultra pure water system D11931, was used throughout. All solvents were degassed by the freeze, pump, thaw method and the water was purged with argon gas for two hours prior to use. Unless otherwise stated, all reactions were carried out utilizing Schlenk line anaerobic techniques under argon.

The Fe/Au nanoparticles were synthesized as previously reported.[16, 18, 19] The preparation was carried out in a reverse micelle system that used cetyltrimethylammonium bromide as the surfactant, octane as the oil phase, and 1-butanol as the co-surfactant. The inner water droplet in the reverse micelle served as a nano-scale reactor for the chemical synthesis. The iron nanoparticles were prepared by the reduction of iron(II) with $NaBH_4$. A reverse micelle aqueous solution of $FeSO_4$ was added to the reverse micelle aqueous solution of $NaBH_4$ in a 1:4 mole ratio and the resulting mixture was stirred at room temperature for one hour. To create a gold shell on the iron core, an aqueous micelle solution containing one mole of $HAuCl_4$ was immediately added to the above solution of $FeSO_4$ and $NaBH_4$. Then 8 moles of the $NaBH_4$ micelle solution was added to the solution and stirred at room temperature overnight. A dark precipitate resulted and was removed using a magnet and subsequently washed twice with methanol to remove any nonmagnetic material present; the product was then dried under vacuum. Several samples were prepared and treated as follows: Fe/Au nanoparticles protected from air-oxidation, **1**, and then exposed to air for one week, **2**, and Fe/Au nanoparticles heated in air for two hours at 870 K, **3**. In all cases, the initial syntheses yield the same X-ray powder diffraction pattern, TEM, magnetization, and chemical analysis. In addition, nanoparticles of iron without a gold shell were prepared following the first part of the scheme outline above, and were protected from air for the subsequent Mössbauer X-ray diffraction and spectral studies. The resulting X-ray powder diffraction pattern was consistent with the presence of α-iron.

Inductively coupled plasma elemental analysis was performed by Desert Analytics, Tempe, AZ, on both the iron nanoparticles prepared without a gold coating and the Fe/Au nanoparticles.

X-ray diffraction measurements were performed on a Scintag PAD-V diffractometer with Cu $K_\alpha$ radiation at a wavelength of 1.5406 Å. The Material Data Inc.

JADE6 software was utilized for data analysis. The x-ray diffraction patterns were collected from 30° to 90° in $2\theta$ with a step size of 0.02° and a dwell time of two seconds. Calculation by Scherrer equation determined the crystallite sizes of nanoparticles: $L = (0.9\lambda)/(\beta\cos\theta)$, where $\lambda$ is the x-ray wavelength in nm, $\beta$ is the intrinsic peak width in radians on a $2\theta$ scale, $\theta$ is the Bragg angle, and 0.9 is the Scherrer constant. The peak at a $2\theta$ of 38.184° or a $\sin\theta/\lambda$ of 0.212 was fit with the Material Data Inc. JADE6 software to determine the crystallite size, after accounting for the instrumental line width.

An additional high-resolution x-ray diffraction pattern was collected on beam line 2-1 at the Stanford Synchrotron Radiation Laboratory. This beam line is equipped[20] with a 2-axis Huber diffractometer and the size of the focused beam is $2 \times 1$ mm$^2$. The x-ray diffraction pattern was collected from 20° to 120° in $2\theta$ at the 9.66 keV Zn K-edge at a wavelength of 1.2838 Å. The sample was placed in a 0.3 mm quartz capillary and sealed with degassed petroleum jelly at the University of California–Davis in order to prevent oxidation; the x-ray diffraction pattern was obtained the next day.

Transmission electron micrographs of the nanoparticles were obtained on a Philips CM-12 transmission electron microscope at 100 keV. A SiO$_2$ grid was dipped in a propanol solution in which the Fe/Au nanoparticles were suspended and the grid was dried in air and then in an oven at 400 K for two hours.

A Quantum Design superconducting quantum interference device (SQUID) magnetometer was used for magnetic measurements. While working under argon to prevent oxidation, 10 mg of sample was placed in a gel-capsule with glass wool and degassed oil and the capsule was then placed in a drinking straw. The temperature and field dependence of the magnetization was measured within an hour after preparation and then after one week.

The first-order reversal curve measurements were performed at 35 K by using a Princeton Measurements vibrating sample magnetometer (VSM) with a liquid helium continuous flow cryostat. For this study, 10 mg of Fe/Au nanoparticles were first dispersed in hexane under sonication inside a glove box, and then mixed with rubber cement. In this technique, the VSM is used to measure a few hundred first-order reversal curves (FORC's) in the following manner. After saturation, the magnetization, $M$, is measured with increasing applied field, $H$, starting from the reversal field, $H_R$, back to positive saturation. A family of FORC's is measured at different $H_R$ values with equal field spacing, thus filling the interior of the major hysteresis loop. The FORC distribution is then defined by a mixed second order derivative,[21-25]

$$\rho(H_R, H) \equiv -\frac{1}{2} \frac{\partial^2 M(H_R, H)}{\partial H_R \partial H}. \tag{1}$$

A two-dimensional contour plot of the distribution, $\rho$, versus $H$ and $H_R$, i.e., a FORC diagram, can then be used to probe the details of the magnetization reversal. Alternatively, $\rho$ can be plotted as a function of the local coercivity, $H_c$, and the bias field, $H_b$, after a $H_b = (H + H_R)/2$ and $H_c = (H - H_R)/2$ coordinate transformation.[21, 22] If a material is composed of a set of independent magnetic particles, the resulting FORC diagram will map the distribution of the coercivity and bias field of the collection of particles. For real systems, the FORC diagram also reveals any complex interactions that may occur among the particles, as will be illustrated below. Thus, the first-order reversal curves provide much more information than the ensemble average measured by typical magnetic major hysteresis loops. Details of the methodology and its applications have been described in prior publications.[21-25]

The Mössbauer spectra were measured at 78 and 295 K on a conventional constant-acceleration spectrometer that utilized a room-temperature rhodium-matrix cobalt-57 source and was calibrated at room temperature with α-iron foil. The studies were performed at these temperatures because of the requirement that the samples not be exposed to oxygen or moisture.[26] Immediately after synthesis the sample was placed in a nitrogen-filled glass vial and sealed; this vial was then placed in a nitrogen-filled plastic vial and shipped to the University of Missouri–Rolla by overnight express mail. The absorber, which was prepared and placed in the cryostat in an inert-atmosphere dry box under pure dry nitrogen, contained ~20 mg/cm$^2$ of material finely dispersed in deoxygenated boron nitride. The hyperfine parameters reported herein have a relative accuracy of ± 0.2 and ± 0.4 T for the $\alpha$-iron field and for the remaining fields, respectively, ± 0.01 and ± 0.02 mm/s for the remaining hyperfine parameters for $\alpha$-iron and the remaining spectral components, respectively, and ± 1 and ± 2 % for the $\alpha$-iron relative area and the remaining spectral component relative areas, respectively. The absolute errors are approximately twice as large.

**Results and Discussion**

We have previously shown that the gold shell on these Fe/Au nanoparticles does not protect the iron containing core from oxidation and we have speculated that this is due to incomplete coverage and cracks in the gold shell.[18, 19] In this paper, we have investigated the composition of the nanoparticles and the oxidation of the iron in the core.

We prepared iron nanoparticles from the reduction of $FeSO_4$ with $NaBH_4$ to produce uncoated "pure" iron nanoparticles for a comparison. We have prepared Fe/Au nanoparticles and provided complete characterization for three samples: **1**, protected from air oxidation, and **2**, subsequently exposed to air for one week, and **3** heated in air at 870 K for 2 hours.

**Elemental Analysis.** Elemental analysis revealed that the uncoated "pure" iron nanoparticles contained 60.7 % iron and 7.5 % boron and, presumably, 31.8 % of other elements. The sample was analyzed only for iron and boron, and the total percentage of identified elements was low. The Fe/Au nanoparticle, **1**, elemental analysis totals were also low with 49.53 % gold, 21.36 % iron, 2.22 % boron, 0.93 % carbon, 3.23 % sulfur, 5.92 % of sodium and, presumably, 16.8 % of other elements. The low total percentage found for the iron nanoparticles is attributed to the presence of salts of $SO_4^{2-}$ that could not be completely removed because of an attempt to maintain an unoxidized product iron core. The total percentage of identified elements observed for the Fe/Au nanoparticles is much better, but may also reflect the presence of salts of $SO_4^{2-}$ that were not completely removed. The mole ratio of Fe:B is 1.6 for the "pure" iron nanoparticles and 1.9 and for the Fe/Au nanoparticles. The amount of boron is higher than expected and, based in part on the Mössbauer spectral results presented below, maybe due to the presence of either $Fe_{1-x}B_x$ and/or $B(OH)_3$ that has not been completely removed during washing. Typically $Fe_{1-x}B_x$ is produced when the $NaBH_4$:Fe ratio is 4:1 and higher,[27] a ratio that was used herein during the iron and gold reduction.

**X-Ray Diffraction.** Powder x-ray diffraction patterns of the synthesized Fe/Au nanoparticles, **1**, see the blue data points in Figure 1, indicate the presence of both crystalline α-iron and gold, which have reflections at very similar 2θ angles, and the absence of any crystalline iron oxides or other possible side products. To determine whether or not any iron oxide forms in the sample upon aging, the Fe/Au nanoparticles were stored in air for one week, **2**, and the powder diffraction pattern was remeasured, see the green data points in Figure 1. The diffraction pattern obtained after one week is similar to the original pattern and reveals no new diffraction peaks. The patterns can be completely indexed with known α-iron and gold reflections; the indices are given in Figure 1. The patterns indicate both that there are no crystalline iron oxides present immediately after synthesis and that any oxidized product is either amorphous or heavily coated with gold. We have previously shown that even heating the sample in air at 670 K overnight does not give rise to any new diffraction peaks that might indicate the presence of iron oxides.[18]

The powder diffraction pattern obtained with synchrotron radiation is very similar to that obtained with the conventional diffractometer but, as expected for a shorter wavelength and a wider $2\theta$ range, additional gold and α-iron peaks are observed, see the red data points in Figure 1. The product is phase pure, based on the synchrotron data.

It should be noted in Figure 1 that all the observed diffraction peaks can be indexed as resulting solely from gold and one might conclude that the x-rays are only probing the gold coating of the particles. However, a calculation of the penetration depth of the Zn $K_\alpha$ radiation at 9.66 keV indicates that this is not the case and the x-rays are probing the entire particle. Thus, we conclude that the powder diffraction patterns are best indexed with both gold and α-iron, a component that is also observed in the Mössbauer spectra discussed below. Further, these results indicate that the other iron containing components observed in the Mössbauer spectra, as well as non-iron containing components, whose presence is indicated by the elemental analyses, must be non-crystalline or at best very poorly crystalline components.

A Scherrer x-ray line width analyis[28] yields an average diameter of 19 nm, a diameter which is consistent with diameters obtained from transmission electron microscopy.

**Transmission Electron Micrographs.** The structure of the Fe/Au nanoparticles may be observed in transmission electron micrographs obtained immediately after synthesis and heating in air at 400 K to prepare the specimen. The results, see Figure 2, confirm the core-shell structure of the nanoparticles which are aggregated on the copper grid because they are magnetic even at room temperature. The interaction of the magnetic fields of the particles with the electron beam also results in a somewhat blurred image. Further, the presence of both iron and gold is confirmed by energy dispersive x-ray spectroscopy. The transmission electron micrograph obtained one week after synthesis is virtually identical with the micrograph shown in Figure 2.

The inset to Figure 2 shows the distribution in diameter obtained for 500 particles. A fit of this distribution with a Gaussian law gives an average diameter of 17 nm with a standard deviation of 4 nm. Based on the TEM micrographs the core of the particles are estimated to range in diameter from 5 to 10 nm. However, aggregation and lack of contrast lead to significant errors in obtaining a more accurate estimate for the core diameter.

**Magnetic Properties.** Both the zero-field cooled and field cooled temperature dependence of the magnetization has been measured in a 100 Oe applied field, see Figure 3, both immediately after synthesis, **1**, and one week later, **2**. The maximum in the zero field cooled curve and divergence in the zero field and field cooled curves indicate a blocking temperature, $T_B$, of at least 300 K. The temperature-dependence of the magnetization remains qualitatively the same after one week, but the magnetization is reduced by a factor of ca. 0.5. This reduction suggests that the nanoparticles are still aggregated after exposure to air, but that there has been some oxidation of the core as the magnetization of iron oxides are lower than α-iron.[29] The diameter of the core of the Fe/Au nanoparticles varies from ~5 to ~10 nm and it would be expected that this range of diameters, and the associated distribution of blocking temperatures, would lead to the broad maximum observed in the field cooled and zero field cooled magnetization.

**Analysis of the First-Order Reversal Curves.** The first-order reversal curves, along with the major hysteresis loops, of a mixture of the Fe/Au nanoparticles with rubber cement have been measured as a function of time in order to monitor the evolution of their magnetic properties. These measurements were carried out after zero-field cooling to 35 K; a nominal field step of 0.02 kOe was used. The magnetization reversal curves filling the interior of the major loop are shown in Figure 4a. The corresponding FORC distribution, $\rho$, is shown in terms of the coercivity, $H_c$, and the bias field, $H_b$, coordinates in Figure 4b. This approach yields a clear indication of any changes in magnetic behavior with time.

As is indicated in Figure 4b, a non-zero FORC distribution $\rho$ extends along the $H_c$ axis, indicating a finite distribution in the local coercivity, a distribution that results from the finite distribution of particle sizes. Furthermore, some subtle changes in the magnetic characteristics of the Fe/Au nanoparticles have been revealed by the FORC measurements. For example, a slow oxidation process of gold-coated iron nanoparticles over time is illustrated in Figure 5. The projection of the FORC distribution onto the $H_c$ axis, see Figure 5a, shows the coercivity distribution as a function of time. The average coercivity (peak position) is reduced with time. For ultrafine single-domain particles, the coercivity is proportional to the particle size.[30, 31] Thus the reduction of $H_c$ is consistent with a gradual decrease in the Fe/Au core size, a decrease that may be assisted by the repeated thermal cycling of the Fe/Au particles from room temperature to 35 K and back, and the concomitant oxidation of the magnetic core as the gold coating is gradually cracked and/or destroyed. Furthermore, the coercivity distribution shown in Figure 5a becomes narrower over time, indicating that particles with larger magnetic cores have experienced more oxidation and enhanced size reduction.

The major loop coercivity and saturation magnetization are also observed to decrease over time, as shown in Figures 5c and 5d. Note that the major loop coercivity shown in Figure 5c is not necessarily the same as the peak coercivity shown in Figure 5a. At 35 K, the major loop coercivity decreases from 170 to 110 Oe over 17 days, with a decay constant of ~42 days required to reach 1/$e$ or 37 %. These coercivities agree well with the 295 K coercivities obtained[27] for α-iron coated with $Fe_{1-x}B_x$. The decreases in the coercivity and saturation magnetization further indicate sample oxidation over time. The decay constant is much longer than that observed if the sample was exposed directly to air.[18] The rubber cement offers some degree of protection from the atmosphere leading to slower oxidation with time than previously reported.[18, 19]

It is also interesting to note the evolution of the distribution of the bias field, $H_b$, a distribution which is related to the average interparticle spacing and the extent of their magnetic interactions.[25] The $H_b$ distribution is obtained by projecting the FORC distribution onto the $H_b$ axis, as is shown in Figure 5b. If the Fe/Au nanoparticles are well dispersed, i.e., if the dipolar and exchange interactions between the particles are negligible, each particle would experience only the applied field and thus reverse its magnetization at its respective coercive field. The hysterons, or hysteresis loops for each particle, would then have zero bias, resulting in a FORC distribution that has a narrow ridge along $H_c$ centered at $H_b = 0$. Because the Fe/Au nanoparticles are not fully dispersed, their interactions are manifested as a distribution of the bias field, $H_b$, as is shown in Figures 4b and 5b. The bias field distribution changes negligibly with time for the Fe/Au nanoparticles, see Figure 5b, indicating that, although the particles are undergoing oxidation, as is indicated by the trends in Figures 5a, c, and d, the average particle spacing and the interactions between the particles remain essentially unchanged.

**Mössbauer Spectral Studies.** The Mössbauer spectra, obtained at 78 and 295 K, of iron nanoparticles prepared both without (uncoated "pure" Fe) and with a gold coating, **1**, are shown in Figures 6 and 7, respectively, and the corresponding hyperfine parameters are given in Table 1. It is obvious from these figures that in both cases the spectra exhibit a sharp sextet, with a relative area of ca. 16 and 40 %, for the iron and Fe/Au nanoparticles, respectively, and with hyperfine parameters that are typical of crystalline α-iron,[26] thus verifying the production, at least in part, of iron particles using the reduction route provided by the reverse micelles. In addition, the spectra of both samples exhibit the presence of paramagnetic high-spin iron(II), high-spin iron(III), and a broad sextet. Finally, the spectra of the iron nanoparticles, see Figure 6, show two broad components with a relative area of ca. 4.5 % and large hyperfine fields, components that can be assigned to poorly crystallized magnetically ordered iron(III) oxides, oxides that most likely result from a slight oxidation of the sample during the preparation of the Mössbauer spectral absorber.

The assignment of the high-spin iron(II) doublet present in the four spectra of Figures 6 and 7 to a specific compound is difficult. Because $FeSO_4$ was used in the preparation, it is tempting to assign this doublet to a ferrous sulfate. However, the observed hyperfine parameters do not match either those of anhydrous $FeSO_4$ or those of $FeSO_4 \cdot 7H_2O$. The hyperfine parameters of the high-spin iron(III) doublet are typical of superparamagnetic particles of $\gamma$-$Fe_2O_3$ or $Fe_3O_4$;[32] it is not possible on the basis of the hyperfine parameters to differentiate these two oxides. This assignment agrees with the presence of $\gamma$-$Fe_2O_3$ in the Fe/Au nanoparticles[17] and the presence of $Fe_3O_4$ in the Fe-Au composite particles[33] observed by x-ray absorption spectroscopy. The iron(III) doublet relative areas of ca. 10 and 22 %, in the iron and Fe/Au nanoparticles, respectively, indicate that only a small fraction of the sample behaves as superparamagnetic particles on the iron-57 Mössbauer-effect timescale of $10^{-8}$ s. This difference with the conclusion drawn from the magnetic measurements is expected because of the very different characteristic measuring times of the two techniques. The small iron(III) oxide particles that exhibit superparamagnetic relaxation on the Mössbauer time scale are smaller than the nanoparticles shown in Figure 2; their typical diameter is estimated to be 10 nm.[32]

The broad sextet with a hyperfine field of ca. 25 T was initially assigned to iron-gold alloys.[19] However, in view of its presence in the Mössbauer spectra of the iron nanoparticles, see Figure 6, this assignment must be revised. The presence of boron revealed by the inductively coupled plasma elemental analysis, as well as the observation of $Fe_2B$ in the iron nanoparticles prepared via a similar route by Glavee et al.,[13] support the presence of $Fe_2B$ or $Fe_{1-x}B_x$ in both the iron and Fe/Au nanoparticles studied herein. The hyperfine parameters of the broad sextet agree reasonably well with those of amorphous $Fe_{1-x}B_x$ alloys,[27, 34] prepared by a method similar to that used herein. The 295 K hyperfine field of 23.4 ± 0.4 T observed herein and the linear dependence[34] of the field with the boron content yield an $x$ value of 0.27 ± 0.02 and an average composition of $Fe_{0.73}B_{0.27}$. Further, the poor crystallinity and/or non-homogeneous composition of the $Fe_{1-x}B_x$ alloy lead to the observed broadened spectral absorption lines observed for this component.

The Fe/Au nanoparticles have been heated at 870 K in air for two hours, **3**, and the Mössbauer spectra of the resulting material have been measured. Any amorphous components present should have oxidized and crystallized and any iron(III) oxide present should become crystalline and thus be detected in the Mössbauer spectra. Figure 8 shows that ordered iron(III) oxide, as well as high-spin, presumably superparamagnetic iron(III), are the main components in the material after heating. The spectra have been fit with two sextets and one doublet. The doublet, which has hyperfine parameters that are similar to the high-spin iron(III) doublet observed in Figures 6 and 7, is similarly assigned to superparamagnetic particles of $\gamma$-$Fe_2O_3$ or $Fe_3O_4$. The sextet with the largest hyperfine field of 50.4 and 53.6 T at 295 and 78 K, respectively, is assigned to ordered iron(III) oxide, possibly $\gamma$-$Fe_2O_3$.[35] The second broad sextet exhibits a hyperfine field that decreases substantially between 78 and 295 K, a decrease that is not expected for the common iron(III) oxides. This decrease indicates that the magnetic ordering temperature of this iron(III) component is not very high. Because of the presence of $Fe_{1-x}B_x$ in the Fe/Au nanoparticles, the presence of an iron boron oxide is quite likely. Even though the observed hyperfine field does not match perfectly with that observed in $FeBO_3$,[36] this sextet is tentatively assigned to $FeBO_3$ or to a related iron borate.

These results are consistent with the suggestion that there is oxidation of the iron core during the exposure to air during heating. It is also possible that any amorphous iron(III) oxide present before heating is crystallized during the high temperature annealing.

## Conclusions

The simplicity of the x-ray diffraction pattern, even using a synchrotron source, strikingly contrasts with the complexity of the Mössbauer spectra. The Mössbauer spectra of both the uncoated and gold coated iron nanoparticles prepared by reduction indicate that three major iron containing components are present. The desired $\alpha$-iron phase represents 16 and 40 percent of the iron components, in the uncoated and gold coated iron nanoparticles, respectively. The other two undesirable components are an amorphous $Fe_{1-x}B_x$ alloy and several poorly crystallized iron oxides species. The simplicity of the x-ray diffraction pattern results from a combination of the accidental overlap of the diffraction peaks for gold and $\alpha$-iron, the broadening of the gold peaks for small particles, and the poor crystallinity or amorphous nature of the two undesirable iron components. A careful examination of the intensity of the peaks at ca. 0.21 and 0.24 in Figure 1 indicates that their intensity ratio is close to 2:1, as would be expected for gold and that there is little, if any, contribution of the most intense diffraction peak for $\alpha$-iron in the peak at 0.24. The magnetic results suggest that aging through thermal cycling of the Fe/Au nanostructured material yields a reduction in size of the magnetic core and a concomitant oxidation.

The ensemble of results presented herein suggests that a number of detailed analyses with different techniques probing different components and properties, are required to fully understand any complex nanoparticles. These results also suggest that further efforts for synthetic optimization of Fe/Au nanoparticles are warranted.

## Acknowledgements

The authors thank Hsiang-Wei Chiu for obtaining the synchrotron x-ray diffraction data, John Neil for technical support during the analysis of the x-ray diffraction results. This research was supported by the National Science Foundation (DMR-0120990, CHE-0210807, and ECS-0508527), the American Chemical Society (PRF-43637-AC10), the Alfred P. Sloan Foundation, and the University of California. Portions of this research were carried out at the Stanford Synchrotron Radiation Laboratory, a national user facility operated by Stanford University on behalf of the U.S. Department of Energy, Office of Basic Energy Sciences. F. G. acknowledges with thanks the financial support of the Fonds National de la Recherche Scientifique, Belgium, through grant 9.456595 and the Ministère de la Région Wallonne for grant RW/115012.

Table 1. Mössbauer Spectral Hyperfine Parameters

| Nano-particles | T, K | H, T | δ, mm/s[a] | QI,[b] mm/s | Γ, mm/s | ΔΓ, mm/s | Area, % | Possible Assignment |
|---|---|---|---|---|---|---|---|---|
| Iron | 296 | 49.5 | 0.297 | 0.01 | 0.94 | 0.00 | 4.5 | ordered iron(III) oxide |
| | | 42.6 | 0.634 | –0.30 | 0.94 | 0.00 | 4.5 | ordered iron(III) oxide |
| | | 32.7 | 0.005 | 0.00 | 0.30 | 0.00 | 16.0 | α-iron |
| | | 23.4 | 0.141 | –0.05 | 0.62 | 0.33 | 44.9 | $Fe_{0.73}B_{0.27}$ |
| | | 0 | 1.033 | 2.19 | 0.61 | - | 17.8 | high-spin iron(II) |
| | | 0 | 0.445 | 0.75 | 0.50 | - | 11.7 | high-spin iron(III) |
| | 78 | 51.3 | 0.385 | 0.00 | 0.52 | 0.00 | 4.4 | ordered iron(III) oxide |
| | | 47.0 | 0.727 | –0.32 | 0.52 | 0.00 | 4.4 | ordered iron(III) oxide |
| | | 33.7 | 0.092 | 0.01 | 0.31 | 0.00 | 16.6 | α-iron |
| | | 25.1 | 0.244 | –0.09 | 0.62 | 0.27 | 45.0 | $Fe_{0.73}B_{0.27}$ |
| | | 0 | 1.151 | 2.44 | 0.68 | - | 20.9 | high-spin iron(II) |
| | | 0 | 0.489 | 0.82 | 0.42 | - | 8.8 | high-spin iron(III) |
| Fe/Au nanoparticles | 295 | 32.8 | 0.014 | 0.00 | 0.31 | 0.00 | 35.8 | α-iron |
| | | 23.4 | 0.226 | –0.20 | 0.488 | 0.00 | 27.0 | $Fe_{0.73}B_{0.27}$ and/or Fe-Au alloy |
| | | 0 | 1.044 | 2.08 | 0.40 | - | 13.3 | high-spin iron(II) |
| | | 0 | 0.255 | 0.75 | 0.69 | - | 24.0 | high-spin iron(III) |
| | 78 | 33.7 | 0.113 | 0.00 | 0.38 | 0.00 | 40.2 | α-iron |
| | | 24.4 | 0.192 | –0.48 | 0.67 | 0.00 | 30.7 | $Fe_{0.73}B_{0.27}$ and/or Fe-Au alloy |
| | | 0 | 1.168 | 2.30 | 0.33 | 0.00 | 9.9 | high-spin iron(II) |
| | | 0 | 0.306 | 0.88 | 0.60 | 0.00 | 19.2 | high-spin iron(III) |
| Fe/Au nanoparticles heated to 870 K | 295 | 50.4 | 0.381 | –0.23 | 0.31 | 0.15 | 39.0 | ordered iron(III) oxide |
| | | 41.5 | 0.411 | 0.25 | 0.50 | 0.61 | 49.2 | ordered iron(III) oxide |
| | | - | 0.332 | 0.86 | 0.94 | - | 11.8 | high-spin iron(III) |
| | 78 | 53.6 | 0.472 | 0.37 | 0.23 | 0.15 | 30.5 | ordered iron(III) oxide |
| | | 50.6 | 0.474 | 0.14 | 0.45 | 0.43 | 63.5 | ordered iron(III) oxide |
| | | - | 0.441 | 0.82 | 0.50 | - | 6.0 | high-spin iron(III) |

[a]The isomer shifts are given relative to room temperature α-iron foil. [b]QI is the quadrupole shifts for the magnetic components and the quadrupole splitting, $ΔE_Q$, for the paramagnetic components.

**Figure Captions**

Figure 1. The powder x-ray diffraction patterns of the Fe/Au nanoparticles obtained immediately after synthesis, **1**, blue, one day after synthesis, red, and one week after synthesis, **2**, green. The blue and green data was obtained with Cu $K_\alpha$ radiation and the red data was obtained with Zn K-alpha synchrotron radiation.

Figure 2. The transmission electron micrograph of the Fe/Au nanoparticles obtained immediately after synthesis, **1**. The inset shows the distribution of Fe/Au nanoparticle diameters.

Figure 3. The temperature dependence of the magnetization of the Fe/Au nanoparticles measured in a 100 Oe applied magnetic field immediately after synthesis, **1**, open circles, and one week after synthesis, **2**, open squares.

Figure 4. A family of first-order reversal curves (FORC) obtained at 35 K four days after the preparation of the Fe/Au nanoparticles, **1**, (a). The outer boundary delineates the major loop. The corresponding FORC distribution is shown in (b).

Figure 5. The projection of the FORC distribution onto the $H_c$, (a), and $H_b$, (b), axes, as well as the decay of major loop coercivity, (c), and saturation magnetic moment, (d), illustrate the time dependence of the oxidation of the Fe/Au

nanoparticles at 35 K. The solid lines in (c) and (d) are fits to exponential decays.

Figure 6. The Mössbauer spectra of iron nanoparticles prepared in a reverse micelle and obtained approximately two weeks after synthesis. The iron(II), iron(III), α-iron, and $Fe_{0.73}B_{0.27}$ components are shown in green, red, black, and blue, respectively.

Figure 7. The Mössbauer spectra of the Fe/Au nanoparticles, **1**, obtained within a week of synthesis, protected from air-oxidation. The iron(II), iron(III), α-iron, and $Fe_{0.73}B_{0.27}$ components are shown in green, red, black, and blue, respectively.

Figure 8. The Mössbauer spectra of the Fe/Au nanoparticles obtained after annealing at 870 K in air for two hours, **3**. The iron(III) components are shown in red.

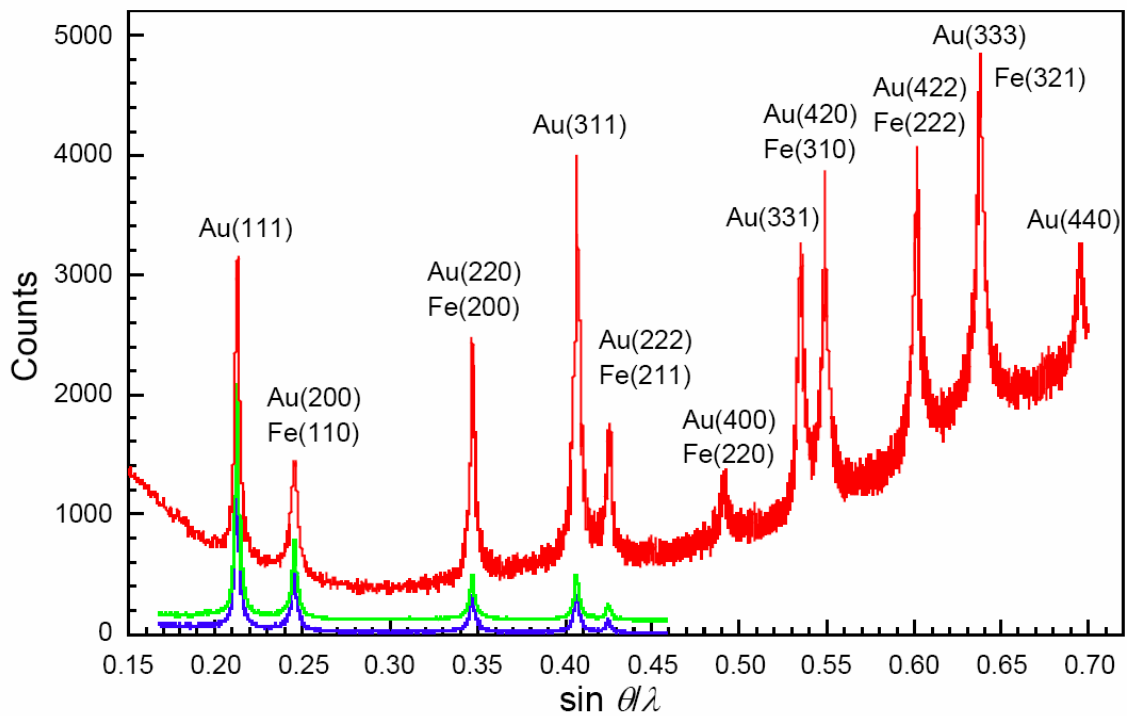

Figure 1.

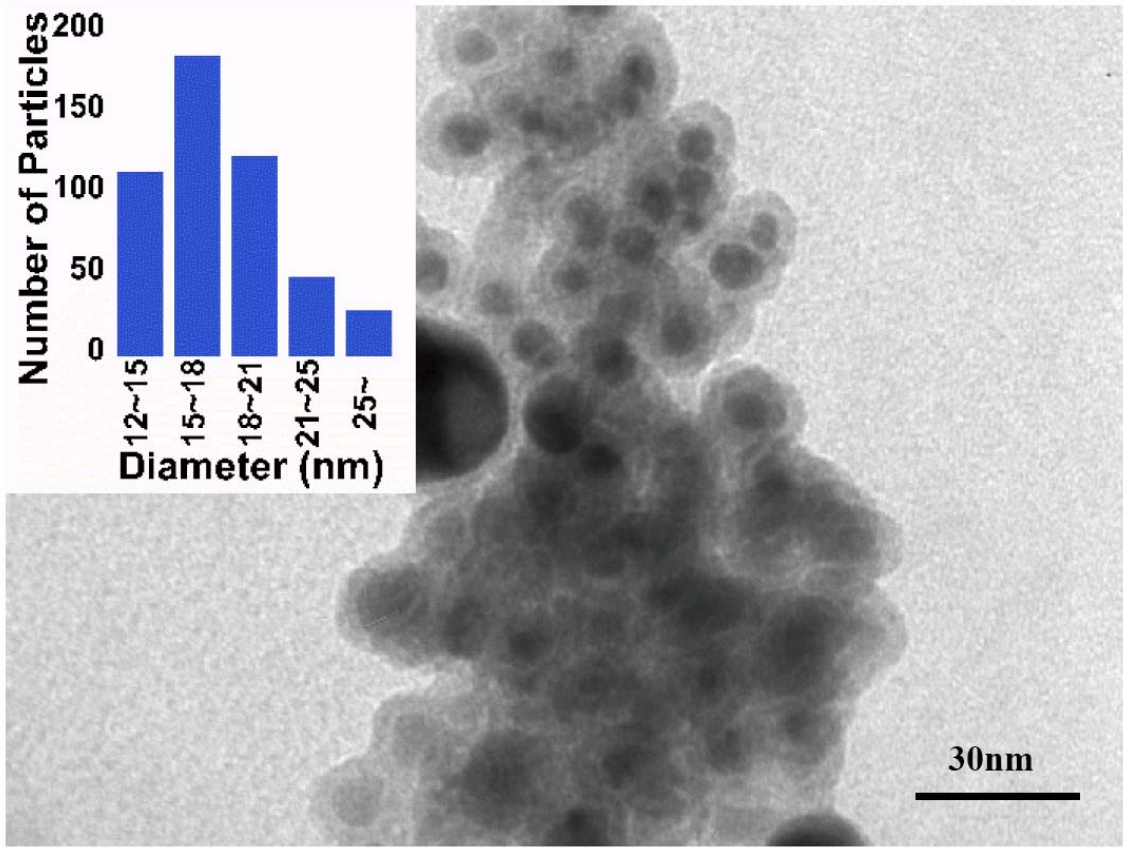

Figure 2.

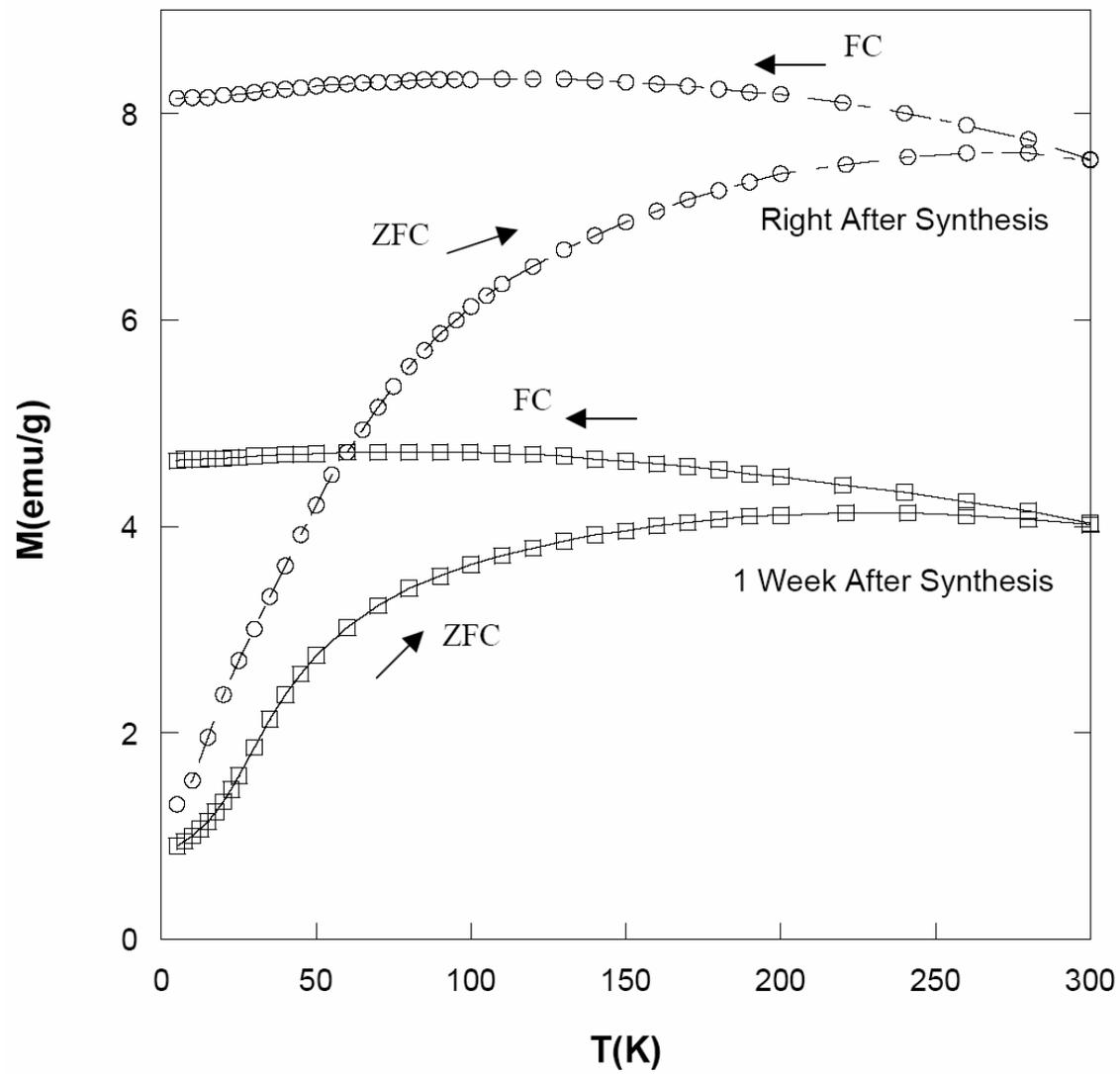

Figure 3.

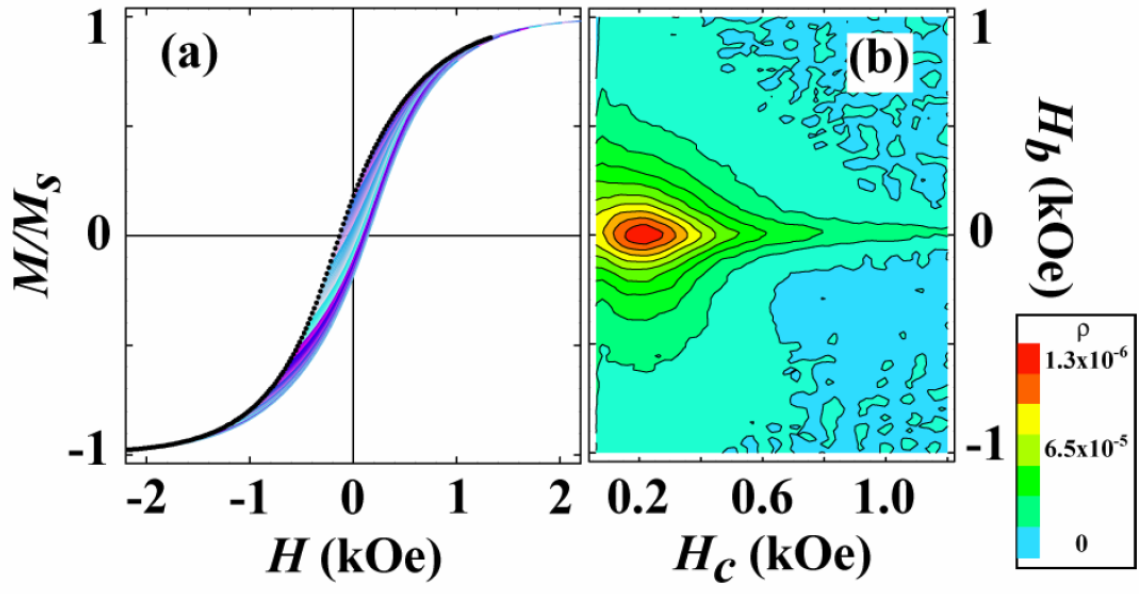

Figure 4.

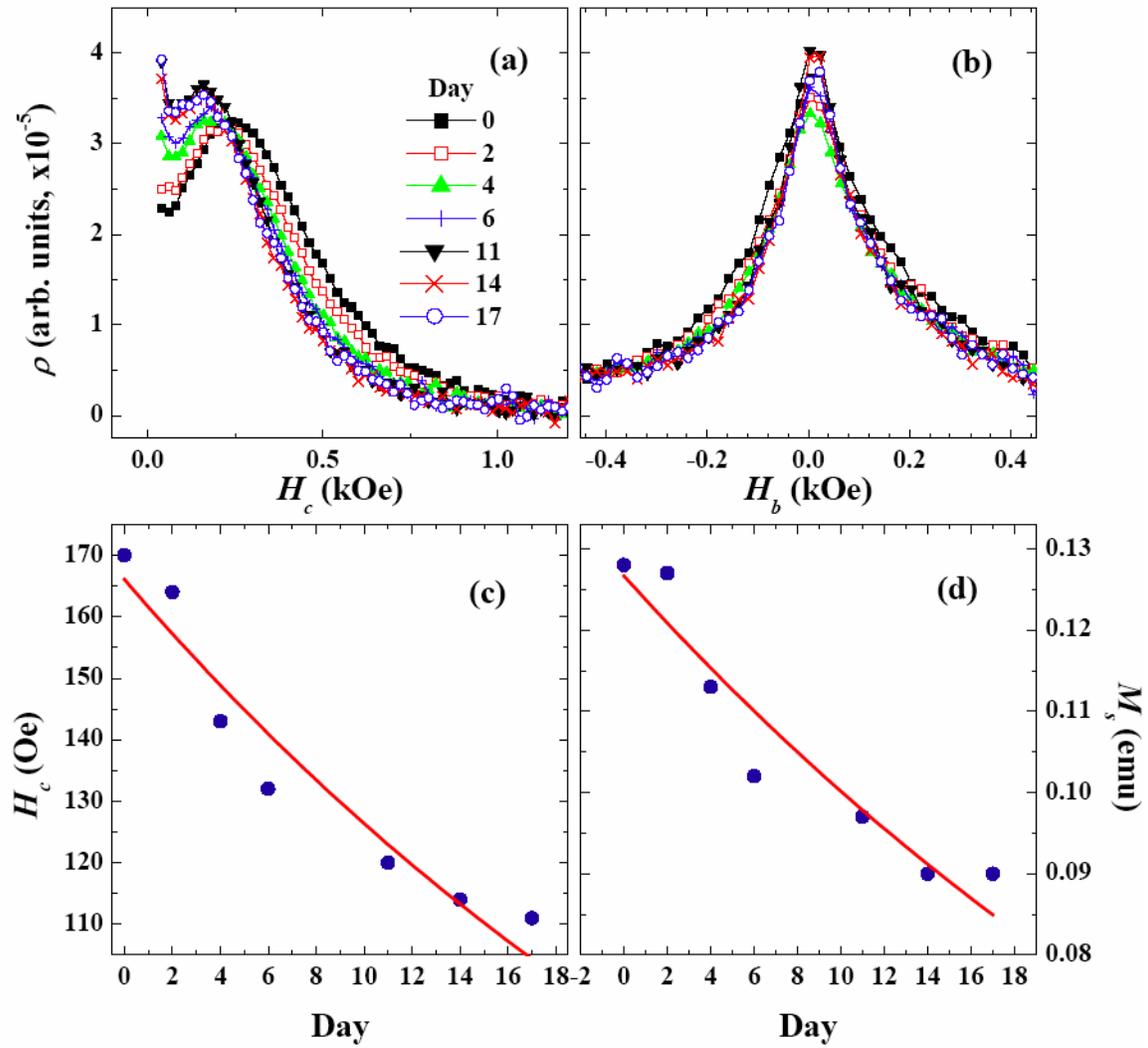

Figure 5.

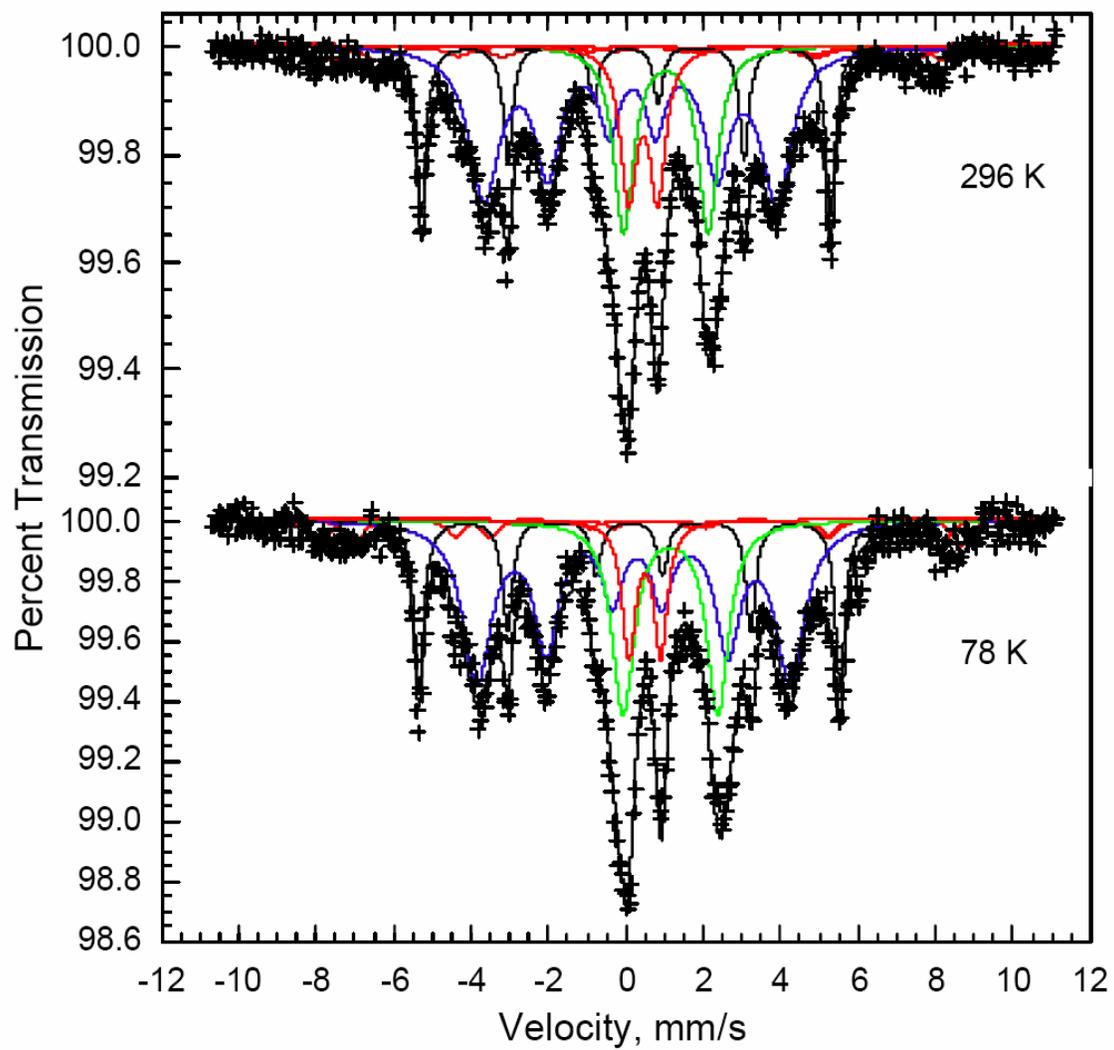

Figure 6.

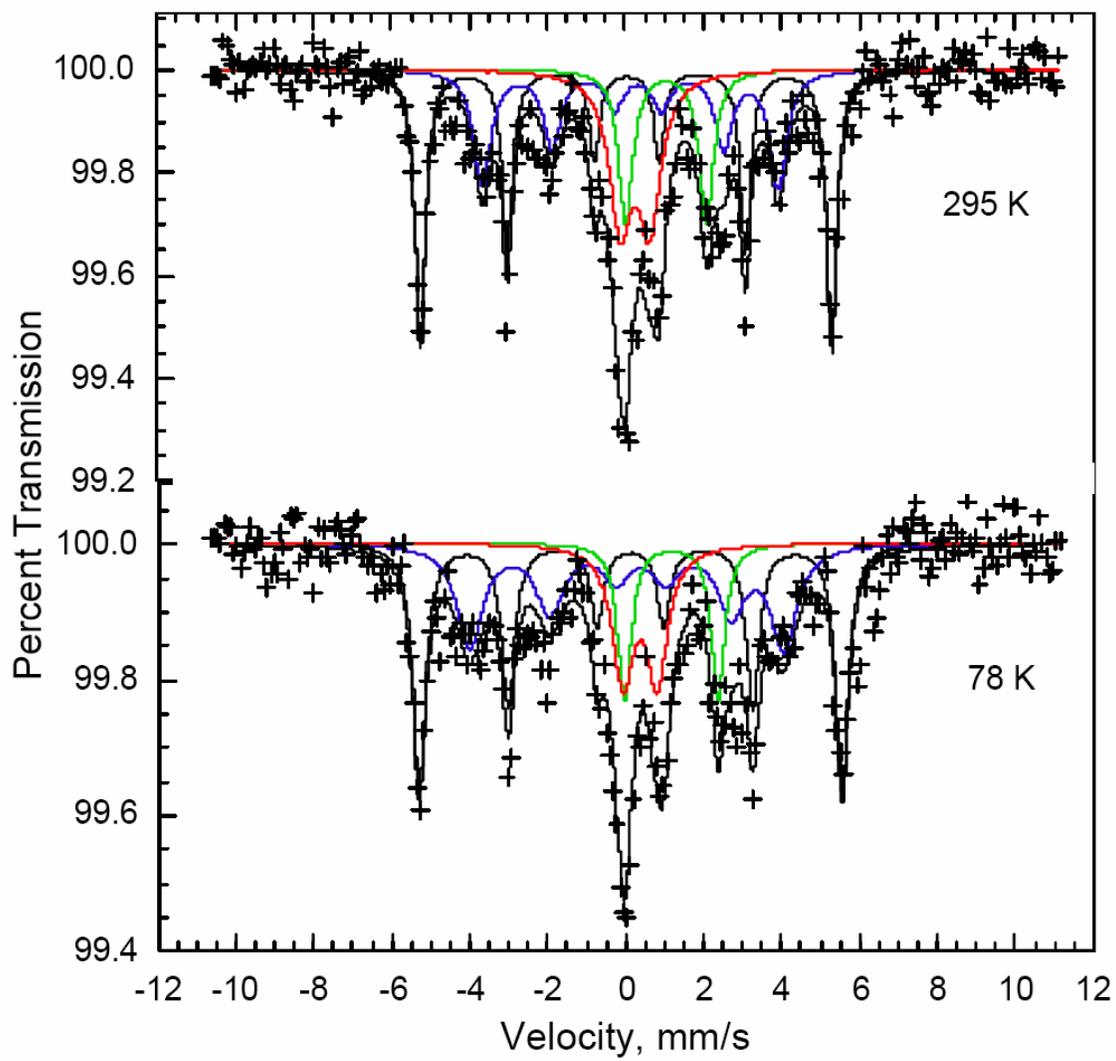

Figure 7.

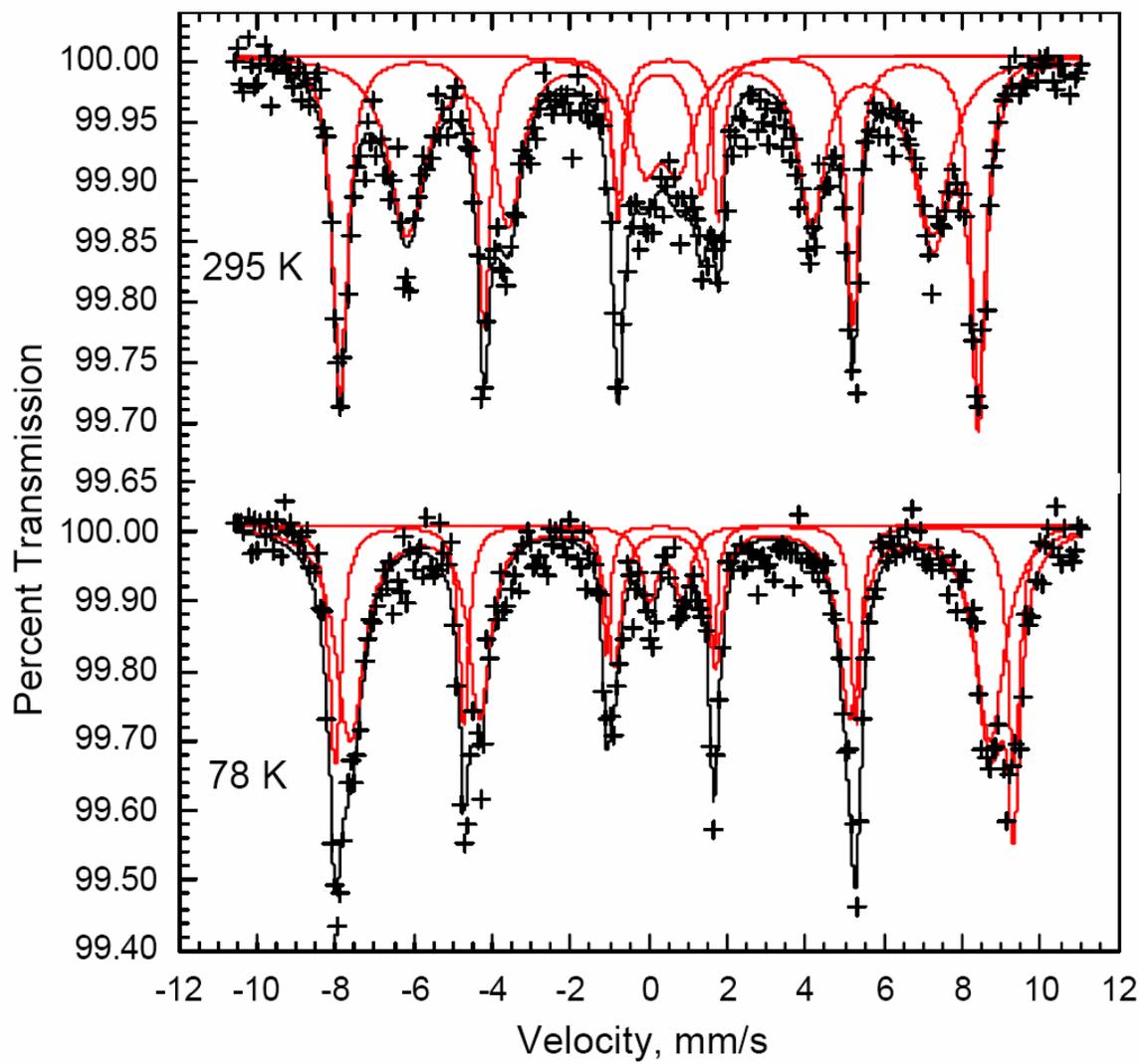

Figure 8.